\documentclass[aps,prl,twocolumn,showpacs,superscriptaddress,floatfix]{revtex4}

\usepackage{graphicx}
\usepackage{epsfig}
\usepackage[english]{babel}

\newcommand{\be}{\begin{equation}}
\newcommand{\bea}{\begin{eqnarray}}
\newcommand{\eea}{\end{eqnarray}}
\newcommand{\ee}{\end{equation}}

\def\one{\ensuremath{\hbox{$\mathrm I$\kern-.6em$\mathrm 1$}}}
\begin{document}

\title{Simulation of many-qubit 
quantum computation with matrix product states}

\author{M. C. \surname{Ba\~nuls}}
\affiliation{Dept. F{\'i}sica Te{\`o}rica and IFIC, U. Val\`encia - CSIC,
46100 Burjassot, Val\`encia, Spain.}
\author{R. \surname{Or\'us}}
\affiliation{Dept. d'Estructura i Constituents de la Mat\`eria,
Univ. Barcelona, 08028, Barcelona, Spain.}
\author{J. I. \surname{Latorre}}
\affiliation{Dept. d'Estructura i Constituents de la Mat\`eria,
Univ. Barcelona, 08028, Barcelona, Spain.}
\author{A. \surname{P\'erez}}
\affiliation{Dept. F{\'i}sica Te{\`o}rica and IFIC, U. Val\`encia - CSIC,
46100 Burjassot, Val\`encia, Spain.}
\author{P. \surname{Ruiz-Femen{\'i}a}}
\affiliation{Max-Planck-Institut f$\ddot{u}$r Physik
(Werner-Heisenberg-Institut),
F$\ddot{o}$hringer Ring 6,
80805 M$\ddot{u}$nchen, Germany.}

%\date{\today}

\begin{abstract}
Matrix product states provide a natural entanglement basis to represent
a quantum register and operate quantum gates on it. This scheme
can be materialized to simulate a quantum adiabatic
algorithm solving hard instances of a NP-Complete problem. Errors
inherent to truncations of the exact action of interacting gates
are controlled by the size of the matrices in the representation.
The property of finding the right solution for an instance and the expected
value of the energy (cost function) are found to be remarkably robust against these errors. 
As a symbolic example, we simulate the algorithm solving a 100-qubit hard
instance, that is, finding the correct product state out of $\sim 10^{30}$
possibilities. Accumulated statistics for up to $60$ qubits seem to point at a sub-exponential
growth of the average minimum time to solve hard instances with
highly-truncated simulations of adiabatic quantum evolution. 
 
\end{abstract}


\maketitle

A detailed understanding of a many-spin quantum system often 
requires its simulation on a classical computer.
Such a possibility is limited to a small number of
spins due to the exponential growth of the size of
the Hilbert space. This is at the heart of the motivation
to build a quantum computer \cite{feynmann}. 
Using standard present technology, a faithful 
simulation of a general Hamiltonian can be achieved for 
systems up to the order of 24 spins. 

Recent developments in representing quantum states
and operating unitary evolution  on them 
have refined the above common lore. 
The idea has evolved from accumulated knowledge
on matrix product states (MPS, related to 
the density matrix renormalization group technique) \cite{mps}
and new insights from quantum information. 
Let us recall that a quantum state for an $n$-qubit system
can be represented by the matrix product 
construction
 \be
\label{state}
 \vert \psi\rangle =
\sum_{\{i\}}\sum_{\{\alpha \}}A^{(1)i_1}_{1\alpha_1}
A^{(2)i_2}_{\alpha_1\alpha_2}
\dots A^{(n)i_n}_{\alpha_{n-1}1}\vert i_1, i_2,\dots,i_n\rangle \ ,
 \ee
where the indices $i_1,\dots,i_n$ for each qubit range from $0$ to $1$ 
(the qubits are placed in a chain)
and $\alpha_1,\dots,\alpha_{n-1}$ are referred to as
ancillae indices that range from 1 to a parameter we shall call $\chi$.
Each matrix $A^{(a)i_a}_{\alpha_{a-1}\alpha_a}$
at site $a$ can be  viewed as a projector from a pair of unphysical ancillae
to the physical degree of freedom that we associate to 
the computational basis. The success of  MPS 
consists in changing the representation
of the quantum state from the computational
basis to a non-local one, closely  attached to entanglement.
To make this comment concrete, let us note that
the matrix representation of a state can be recovered
{\sl via} a chain of Schmidt decompositions that separate
a local system at a time, as made explicit by Vidal \cite{vidal}. More specifically,
$A^{(a)i_a}_{\alpha_{a-1}\alpha_a} = \Gamma^{(a)i_a}_{\alpha_{a-1}\alpha_a}
\lambda^{(a)}_{\alpha_a}$, $\lambda^{(a)}_{\alpha_a}$ being the Schmidt
coefficients of the cut of the system between the $a$ and $a+1$
sides, and $\Gamma^{(a)}$ being tensors for qubit $a$. 
The larger the entanglement is 
for different partitions of the system, the larger is the
needed ancillae space, which corresponds to a higher rank $\chi$. 
MPS can handle
simulations of various dynamics of spin chains with up to hundreds of spins
because their little amount of entanglement
can be represented with $\chi = O({\rm poly}(n))$ \cite{vidal, spinchain}. 
A number of new developments have popped
 up from the basic MPS in the context of quantum information. 
In ref. \cite{vidal}, an efficient
implementation of Hamiltonian evolution was constructed
for slightly entangled systems. An explicit renormalization
group transformation on quantum states was made
explicit using MPS \cite{RG}. The rigid linear structure
of MPS is being now abandoned in favor of 
the more general projected entangled-pair states (PEPS) that have been successfully applied to 
higher dimensional systems \cite{cirac2d}. 

The natural question arises of whether MPS can
be applied to simulate a quantum computer. The content of
this paper is aimed to show that this is indeed possible and
that we can handle large simulations with controlled 
accuracy.
As we shall describe, 
each time an entangling gate is operated on two neighboring
qubits, the range of the connected ancillae index is doubled. 
This is the way interacting gates entangle the system.
To keep the simulation under control, a 
(non-unique) truncation scheme is needed that stops the 
exponential growth of ancillae dimensions. We expect this approximation scheme
to fail whenever the inherently needed $\chi$ is $O(2^n)$. 
Nevertheless, in some of these cases keeping $\chi =
O({\rm poly}(n))$ in the simulation already gives reasonable
approximations to the exact calculation, as we shall see.     

Our presentation will be made
concrete by showing an MPS simulation of quantum computation
in the case of adiabatic evolution for the
NP-Complete Exact Cover satisfiability problem \cite{farhi, gareyjohnson}.
An instance of Exact Cover is defined by a set of $m$
3-bit clauses with satisfying assignments
001, 010 or 100. The problem is defined 
as deciding whether a given instance
accepts a global satisfying assignment of $n$ bits.
This satisfiability problem is NP-Complete.
Classically hard instances of Exact Cover seem to appear
at the so-called easy-hard-easy transition around 
$m\sim .8 n$ \cite{transition}. We have constructed such hard instances, with the additional 
property of having a unique satisfying assignment. The generation of 
hard instances is in itself a difficult problem for which we have developed specific
algorithms, all of them based on the iterative addition of random 
clauses that strictly decrease the number
of solutions of the instance until a single satisfying assignment is
reached.  

The quantum algorithm for a given Exact Cover instance
follows the adiabatic evolution
of the ground state of a Hamiltonian (cost operator) defined by $H(s)=(1-s) H_0 + s H_P$,
where the adiabatic parameter is $s=t/T$ and $t$ runs up to
a total predetermined time $T$. We take the
initial Hamiltonian to be $H_0=\sum_{i=1}^n \frac{d_i}{2} (1-\sigma_i^x)$
where $d_i$ stands for the number of clauses
qubit $i$ enters. The non-local 
problem Hamiltonian corresponds to
the sum of clauses defined as $H_P=\sum_{c(i,j,k)} (z_i+z_j+z_k-1)^2$
where $z_i=(1-\sigma^z_i)/2$ has eigenvalues 0 and 1, and $c(i,j,k)$ stands
for a clause involving qubits $i$, $j$ and $k$.
Exact simulations of quantum algorithms by adiabatic evolution solving hard
instances of satisfiability problems have been carried
so far up to 30 qubits \cite{hogg}. The explosion of entanglement 
between random cuts in the quantum register was first analyzed in ref.
\cite{orus}. The adiabatic evolution drives the system
near a quantum phase transition at $s\sim .69$
following universal scaling laws. Entropy for
half-cuts of the register approximates
on average the scaling law  $S\sim .1 n$, which almost saturates the
maximum $S=n/2$.
This implies that the quantum algorithm cannot
be simulated efficiently in a classical computer \cite{vidal}.
Yet, the fact that entropy does not reach its allowed
maximum suggests that an adequate handling of entanglement
may provide a way to extend simulations far from
naive limitations.

Let us now turn to discuss the detailed way MPS can handle 
the simulation of the adiabatic evolution of Exact
Cover. The simulation
needs to follow a time evolution controlled by 
the $s$-dependent Hamiltonian. This continuous
unitary time evolution can be discretized as follows: 
$U_{T,0}=U_{T,T-\Delta}\dots U_{2\Delta,\Delta} U_{\Delta,0}$
where the increment 
$\Delta\equiv \frac{T}{M}$ defines the discretization, $M$ being a
positive integer. Our simulations indicate that we can take $\Delta=0.125$
while keeping sufficient accuracy (as compared to smaller $\Delta$) 
in all of them. We have explicitely checked that simulations performed  
with $\Delta < 0.125$ lead to equally-good discretizations of the
continuous-time adiabatic algorithm, in the sense that the obtained 
results do not practically differ from the ones calculated for 
$\Delta=0.125$. After $l$ steps 
$s=\frac{t}{T}=\frac{l \Delta}{T}=\frac{l}{M}$,
being $l=0,\dots M$. At any point in the
evolution Trotter's formula to second order is used to divide the unitary
operation $U_{(l+1)\Delta,l \Delta}$ into elementary gates: 
$U_{(l+1)\Delta,l \Delta}=e^{i \Delta H(s)} \sim \left(
e^{i \frac{\delta}{2} (1-s) H_0} e^{i \delta 
s H_P} e^{i\frac{\delta}{2} (1-s) H_0}\right)^{\Delta\over \delta}$.
We have verified that we can maintain a faithful simulation
with $\delta=\Delta$.
The split of exponentials in Trotter's expansion
 is chosen so that $H_0$
is separated from $H_P$. This brings the advantage
that each piece of 
the evolution operator can be decomposed in 
mutually commuting elementary gates: 
\be
\label{h0diag}
e^{i \frac{\delta}{2}(1-s) H_0}=\prod_{i=1}^n e^{i \frac{\delta}{4}
(1-s) d_i(1-\sigma_i^x)}\ ,
\ee
and
\bea
\label{hpdiag}
\nonumber
e^{i \delta s H_P}=&\prod_{c(i,j,k)} e^{i \delta s (z_i+z_j+z_k-1)^2}\\
\nonumber
=&\prod_{c(i,j,k)} e^{i \delta s (z_i^2-2 z_i)} e^{i \delta s (z_j^2-2 z_j)}
 e^{i \delta s (z_k^2-2 z_k)}e^{i \delta s}\\
&e^{i 2\delta s z_i z_j}
e^{i 2\delta s z_i z_k}e^{i 2\delta s z_j z_k }\ .
\eea
The adiabatic evolution
is thus finally reduced to a series of one and two-qubit gates. 
The detailed way these gates operate
on the MPS follows the original
idea of ref. \cite{vidal}:
%\begin{enumerate}
%\item  

\vspace{6pt} 
1. A one-qubit gate acting on qubit $a$ 
only involves an updating of  $A^{(a)}$ that goes as follows:
\be
\label{one-qubit}
U^{(a)} A^{(a)i_a}_{\alpha\beta} \vert i_a\rangle=
A^{(a)i_a}_{\alpha\beta} U^{(a)}_{i_a i'_a}\vert i'_a\rangle \ ,
\ee
which corresponds to the local updating rule
\be
\label{one-qubit-update}
{A'}^{(a)i'_a}_{\alpha\beta}=U^{(a)}_{i_ai'_a}A^{(a)i_a}_{\alpha\beta} \ .
\ee
This gate does not affect ancillae indices. Entanglement
is unaffected as we are just performing local
operations. 

As an example, consider the one-qubit gate $U^{(a)} =
\sigma_x^{(a)}$, $\sigma_x^{(a)}$ being the usual Pauli matrix 
\be
\sigma_x^{(a)} = 
\left( \begin{array}{cc}
  0  &  1 \\
  1  &  0 \\
\end{array}\right) 
\ee
acting on qubit $a$. Then, we have the following simple updating rule for
$A^{(a)}$: 
\be
\left(\begin{array}{c}
{A'}^{(a) 0}_{\alpha\beta} \\
{A'}^{(a) 1}_{\alpha\beta} \\
\end{array}\right)
=  
\left(\begin{array}{cc}
  0  &  1 \\
  1  &  0 \\
\end{array}\right) 
\left(\begin{array}{c}
{A}^{(a) 0}_{\alpha\beta} \\
{A}^{(a) 1}_{\alpha\beta} \\
\end{array}\right)
= 
\left(\begin{array}{c}
{A}^{(a) 1}_{\alpha\beta} \\
{A}^{(a) 0}_{\alpha\beta} \\
\end{array}\right)
\ .
\ee

\vspace{6pt}
%\item  
2. A two-qubit gate involving contiguous qubits $a$ and $a+1$
follows a similar strategy. Let us define 
\be
\label{two-qubit}
U^{(a,a+1)}_{i'_a i'_{a+1},i_a i_{a+1}}
A^{(a)i_a}_{\alpha\beta}
A^{(a+1)i_{a+1}}_{\beta\gamma}\equiv
\Theta^{i'_a i'_{a+1}}_{\alpha\gamma} \ . 
\label{tet}
\ee
Unlike one-qubit gates, interacting gates do not preserve the product
form of the tensors $A$. To reestablish
the MPS structure we need to rewrite $\Theta$
using a Schmidt decomposition. The procedure to follow
is to compute the reduced density matrix from the cut of the system
between the $a$ and $a+1$ sites, which for the right side reads 
$\rho^{ij}_{\alpha\gamma}=|\lambda^{(a-1)}_{\beta}|^2 {\Theta}^{k
  i}_{\beta\alpha}{\Theta^*}^{k j}_{\beta\gamma}$, 
where we have made use of the
$\chi$ known Schmidt coefficients $\lambda^{(a-1)}_{\beta}$ for the
cut between the $a-1$ and the $a$ sites. After the diagonalization of 
$\rho$ using
$(i\alpha)$ and $(j\gamma)$ as composed indices, we directly read from the
eigenvalues the updated $2\chi$ Schmidt coefficients ${\lambda '}^{(a)}_{\beta}$ for this
cut, and the updated matrices ${A'}^{(a+1)i_{a+1}}_{\beta \gamma}$ 
from the coefficients of the eigenvectors. Finally, the new tensors for qubit
$a$ are easily calculated as ${A'}^{(a)i_a}_{\alpha\beta} =
{A'}^{(a+1)i_{a+1}}_{\beta \gamma} \Theta^{i_a i_{a+1}}_{\alpha \gamma}$. 

Let us clarify this procedure with a simple example: consider the quantum state of
two qubits 
\be
|\psi \rangle = |00\rangle \ .
\ee
It is easy to verify that the above state is described by the following values
of the matrices $A^{(a)}$:
\begin{eqnarray} 
A^{(1) 0}_{1 \alpha} = A^{(2) 0}_{\alpha 1} &=& \delta_{1,\alpha} \nonumber \\
A^{(1) 1}_{1 \alpha} = A^{(2) 1}_{\alpha 1} &=& 0  \ . 
\end{eqnarray}
Notice that since the state is separable $\chi = 1$. At this point, let us
apply the two-qubit gate 
\be
U^{(1,2)} = 
\left(\begin{array}{cccc}
\frac{1}{\sqrt{2}} & 0 & 0 & \frac{1}{\sqrt{2}} \\
0 & 1 & 0 & 0 \\
0 & 0 & 1 & 0 \\
\frac{1}{\sqrt{2}} & 0 & 0 & -\frac{1}{\sqrt{2}} \\
\end{array}\right)  
\ee
to the quantum state $|\psi\rangle$: 
\be
U^{(1,2)}|\psi\rangle = \frac{1}{\sqrt{2}} \left(|00\rangle +
|11\rangle\right) \ . 
\ee
Since the resultant state is a maximaly entangled state of $2$ qubits, we
expect $\chi$ to be bigger than $1$. In order to evaluate the updated 
matrices ${A'}^{(a)}$ for qubits $1$ and $2$
we compute the quantity defined in equation (\ref{tet}), which in our case
turns out to be
\begin{eqnarray}
\Theta^{00} = \Theta^{11} &=& \frac{1}{\sqrt{2}} \nonumber \\
\Theta^{01} = \Theta^{10} &=& \frac{1}{\sqrt{2}} \ .
\end{eqnarray}
The density matrix for qubit $2$ (which in this case is equivalent to the
density matrix for qubit $1$) then reads
\be
\rho = 
\left(\begin{array}{cc}
\frac{1}{2} & 0 \\
0 & \frac{1}{2} \\
\end{array}\right)  \ .
\ee
Since the above density matrix is already diagonal, it is clear that the
updated Schmidt coefficients will be 
\be
{\lambda'}^{(1)}_{1} = {\lambda'}^{(1)}_{2} = \frac{1}{\sqrt{2}} 
\ee
and, as expected, we see that $\chi = 2$ since entanglement has been created by the
two-qubit gate. From the above expressions it is simple to get the value of
the updated matrices ${A'}^{(a)}$ for qubits $1$ and $2$: 
\begin{eqnarray}
{A'}^{(1)0}_{11} = {A'}^{(2)0}_{11} &=& 1 \nonumber \\
{A'}^{(1)1}_{12} = {A'}^{(2)1}_{21} &=& 1 \nonumber \\
{A'}^{(1)0}_{12} = {A'}^{(2)0}_{21} &=& 0 \nonumber \\
{A'}^{(1)1}_{11} = {A'}^{(2)1}_{11} &=& 0 \ .  
\end{eqnarray}

\vspace{6pt} 
%\item  
3. Operations involving non-contiguous qubits (as in Exact Cover clauses)
can be reduced to the case 2 using SWAP operations, producing 
an overhead of $O(n)$ operations per clause.

%\end{enumerate}
\vspace{6pt}

The exact simulation of a quantum computer is then completely
defined. The running time of this algorithm scales as
$\sim T n m \chi^3$.
Efficiency depends on the way the growth of the ancillae
space is handled. 
To keep the simulation under
control we define a truncation scheme of the exact simulation.
We choose to use a local procedure, namely, we
keep the first $\chi$ terms out of the $2\chi$ in the Schmidt decomposition
defined in the point 2 above. Only the terms that carry most of the
entanglement in the decomposition are kept \cite{vidal}. This 
reasonable truncation 
carries an inherent -but always under control- loss of unitarity, since
the sum of the retained squared eigenvalues will not
reach $1$. As we shall see, larger $\chi$'s allow
for more faithful simulations. Alternatively, it would be possible to recast 
the whole
enlarged state into its original size in an 
optimal way \cite{cirac2d}. While this second method is manifestly 
more precise, it carries an operational time overhead. It is then 
worth analyzing both techniques. In this paper we shall focus on 
the first one and leave the second for a separate
publication.

We have implemented a number of optimizations 
upon the above basic scheme. 
For any non-local gate there is
an overhead of SWAP operations that damage the 
precision of the computation. To minimize this effect,
every three-qubit clause is operated as follows: we 
bring together the three qubits with SWAPs of the 
left and right qubits keeping the central one 
fixed and, then, we operate  
the two-qubit gates. Before returning the qubits
to their original position we check if any
of them is needed in the next gate. If so,
we save whatever SWAP may be compensated between the
two gates. Ordering of gates is also used to
produce a saving of $\sim 2/3$ of the naive SWAPs. 
Diagonalization of the density matrix in 
the minimum allowed Hilbert space  
is used as well. A further improvement is to keep
a dynamical and local $\chi$, so that ancillae indices
at the different partitions are allowed to take independent values and
grow up to site-dependent limits. This procedure, though, 
has shown essentially no improvement upon a fixed-$\chi$ strategy.

\begin{figure}[t]
  %\centering
  \includegraphics[angle=270,width=1.0\linewidth]{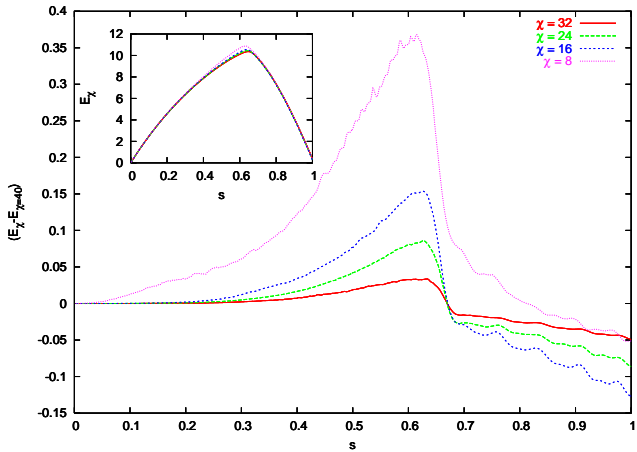}
  \caption{(Color online) Computation of the absolute error (compared to the $\chi = 40$ case) of
 the expected value of the energy (cost function)
along the adiabatic evolution for a typical instance with
30 qubits and 24 clauses for $T=100$ as $\chi$ increases. 
Note the increasing precision with larger $\chi$ as $s$ approaches the phase
 transition from the left-hand-side. 
In the inset, the absolute cost function is
plotted. A similar behavior is also obtained for other instances, all of them 
getting the exact solution at the end of the computation.}
  \label{FigEnergy}
\end{figure}

Let us now focus on the results. We first simulate the
adiabatic algorithm with the requirement that the right solution is found 
for a typical instance of $n=30$ qubits with $m=24$ clauses and 
$T=100$. Along the evolution we compute the expected value of the energy (cost function)
of the system, which can be calculated in $O(n \ {\rm poly(\chi)})$ time. 
This is shown in Fig. \ref{FigEnergy}. The system remains remarkably 
close to the instantaneous ground-state cost function along the approximated evolution.
\begin{figure}[t]
  %\centering
  \includegraphics[angle=270,width=1.0\linewidth]{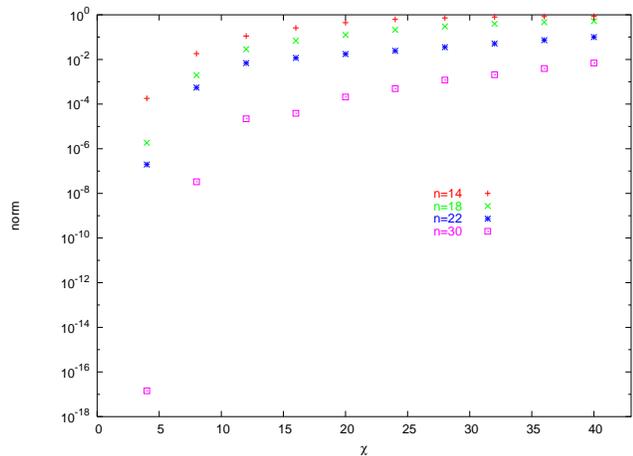}
  \caption{(Color online) Norm in the register as a function of $\chi$ in logarithmic
 scale, for instances of $14,18,22$ and $30$ qubits.}
  \label{FigUnitarity}
\end{figure}
The error in the cost function is minimized as $\chi$ increases.
It is noteworthy to observe how the error in the simulation of the adiabatic algorithm
peakes at the phase transition point. We have also checked that it is precisely at this
point where each qubit makes a decision towards its final
value in the solution. Physically, the algorithm builds entanglement
up to the critical point where the solution is singled out
and, thereon, the evolution drops the superposition of wrong states 
in the register.

This success comes at the price of a controlled loss of unitarity.
We plot in Fig. \ref{FigUnitarity} the norm in the simulation as
a function of $\chi$ in logarithmic scale, for instances of $14,18,22$ and
$30$ qubits. The remarkable fact is that some observables, like
the energy, appear to be
very robust against this inaccuracy. Our simulations also allow to compute the
decay of the $\chi$ Schmidt coefficients $\lambda^{(a)}_{\alpha}$
at any step of the computation. Close to criticality, and for the central
cut of the system, these can be approximately fitted by the law 
$\log_2 (\lambda^{(n/2)}_{\alpha}) = b + \frac{c}{\sqrt{\alpha}} + 
d\sqrt{\alpha}$, with appropriate coefficients $b,c$ and $d$.  

\begin{figure}[t]
  %\centering
  \includegraphics[angle=270,width=1.0\linewidth]{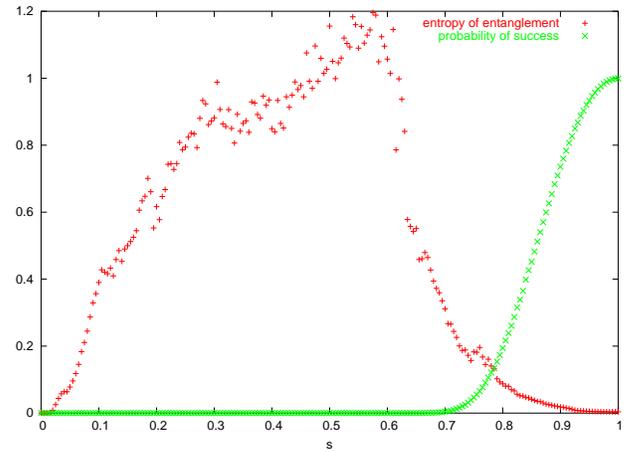}
  \caption{(Color online) An instance with $n=100$ qubits and $m=84$ clauses
is solved using adiabatic evolution with $\chi=14$. The plot
shows the entanglement entropy of a half-cut and the probability in solution state
vs. $s$.}
  \label{Fign100}
\end{figure}

The ultimate goal of finding the correct solution 
appears also to be very robust in the simulations we have performed. The exact
probability of success can be calculated in $O(n \ {\rm poly}(\chi))$ time as
well. As a symbolic example, our program has solved an instance with
$n=100$ qubits, that is, the adiabatic evolution algorithm
has found the correct product state out of $2^{100} \sim 10^{30}$ for
a hard instance with $m=84$ clauses and $T=2000$. The simulation
was done with a remarkable small $\chi=14 \ll 2^{50} = \chi_{max}$ and
is presented in Fig. \ref{Fign100}.
\begin{figure}[t]
  %\centering
  \includegraphics[angle=270,width=1.0\linewidth]{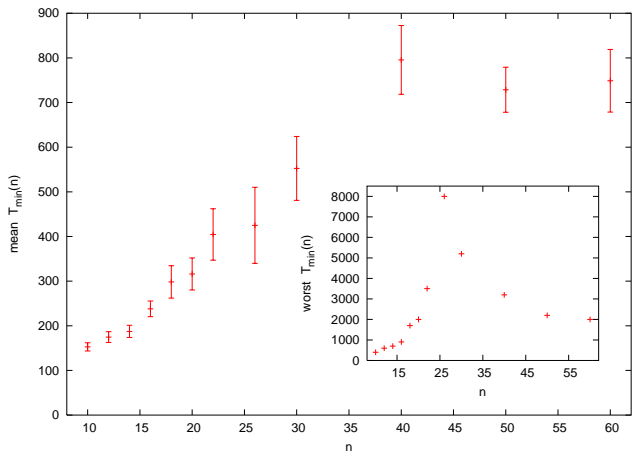}
  \caption{(Color online) Accumulated statistics up to $n=60$ for $T_{min}(n)$ for which
each instance is solved (mean and worst cases). Averages are performed over $200$ instances
for each $n$, except for $n=50,60$ with respectively $199,117$ instances. 
Error bars give $95$ per cent of confidence level in the mean. 
The worst cases found are shown in the inset.}
  \label{FigStat}
\end{figure}

The same robustness of evolving towards the correct solution is 
found for any number of qubits
and small $\chi$. We 
have launched a search for the minimal $T$ that solves samples
of $n$-qubit hard instances in the following way: for a set of small values of
$\chi$, we try a random instance with an initial {\sl e.g.} $T=100$. 
If the solution is
found, we proceed to a new instance, and if not, we restart
with a slower adiabatic evolution {\sl e.g.} $T=200$. This
slowing down of the algorithm
is performed till a correct solution is found and
the minimal successful $T_{min}$ is stored. 
Our results are shown in Fig. \ref{FigStat}. 
The average over $n$-qubit instances appears
to grow sub-exponentially with $n$. In fact, a quadratic 
fit reproduces the data for $n<22$,
consistently with the results found in \cite{farhi}.
The required times for larger
$n$ lie below the extrapolated curve. Isolated instances, however,
may require larger times. Since
the worst $T_{min}$ found depends on the interpolating path, finding an
instance that needs a very large $T_{min}$ is no counterproof for the
efficiency of the adiabatic algorithm, as alternative paths may solve the
instance in a shorter $T$ \cite{farhi}. 

In this paper we have presented
simulations of quantum computation 
based  on matrix product states that can be taken
up to 100 qubits. The remarkable fact that the algorithm finds
the correct solution to a large 
hard instance and the robustness in the expected energy 
is to be contrasted with the loss of unitarity 
inherent to the local truncation scheme.
This drawback may well be ameliorated if optimal 
truncations are implemented.

{\bf Acknowledgments:} we acknowledge discussions with
I. Cirac, E. Farhi and G. Vidal, and support from  
FPA2001-3598, GC2001SGR-00065, GVA2005-264 and FPA2002-00612. 
We would like to express our gratitude for the
use of the GRID computing resources (GoG farm) and the support of
computing technical staff of IFIC.

\end{document}